\documentclass[sigconf]{acmart}
\AtBeginDocument{%
  }

\setcopyright{acmlicensed}
\copyrightyear{2024}
\acmYear{2024}
\acmDOI{XXXXXXX.XXXXXXX}

\acmConference[ACM]{ACM International Conference}{ACM}{ACM}
\begin{document}

\title{Thinking beyond Bias: Analyzing Multifaceted Impacts and Implications of AI on Gendered Labour}

\author{Satyam Mohla}
\affiliation{%
  \institution{Digital Asia Lab, India}
    \country{}
}
\affiliation{%
  \institution{IIT Bombay, India}
\country{}
}
\email{satyammohla@gmail.com}
\orcid{0000-0002-5400-1127}
\author{Bishnupriya Bagh}
\affiliation{%
  \institution{IIT Bombay, India}
  \country{}}
%\country{Austria}
\author{Anupam Guha}
\affiliation{%
 \institution{IIT Bombay, India}
 \country{}}
\renewcommand{\shortauthors}{Mohla, et al.}

\begin{abstract}
Artificial Intelligence (AI), with its multifaceted technologies and integral role in global production, significantly impacts gender dynamics, particularly in gendered labor. This paper emphasizes the need to explore AI’s broader impacts on gendered labor beyond its current emphasis on the generation and perpetuation of epistemic biases. We draw attention to how the AI industry, as an integral component of the larger economic structure, is transforming the nature of work. It is expanding the prevalence of platform-based work models and exacerbating job insecurity, particularly for women. Of critical concern is the increasing exclusion of women from meaningful engagement in the digital labor force. This issue, often overlooked, demands urgent attention from the AI research community. Understanding AI’s multifaceted role in gendered labor requires a nuanced examination of economic transformation and its implications for gender equity. By shedding light on these intersections, this paper aims to stimulate in-depth discussions and catalyze targeted actions aimed at mitigating the gender disparities accentuated by AI-driven transformations.
\end{abstract}




\received{20 February 2007}
\received[revised]{12 March 2009}
\received[accepted]{5 June 2009}

\maketitle

\section{A Critique of the Critique}
Throughout the history of technological development, its impact on gender has often been overlooked, and it has erroneously been considered both gender-neutral and apolitical. This oversight has led to structural issues in research \cite{wajcman2000reflections}. Even when the problem was recognized, efforts primarily focused on addressing bias and increasing diversity within existing technological frameworks, rather than delving into the root causes of how these structures affect gendered labor \cite{adam1996constructions}.

Within the tech sector, gender roles have become so deeply ingrained that substantial gender imbalances in education and workplaces, along with the under representation of women and non-binary individuals in leadership roles, have become widespread and conspicuous issues \cite{cockburn1993gender}. The challenge of women leaving the STEM academic pipeline took time to be acknowledged \cite{london2011influences}. Once recognized, this issue was primarily critiqued in terms of bias—both the epistemic bias in knowledge production due to the individuals creating these technologies \cite{adam1995feminist} and the inherent bias in the artifacts produced by the AI industry. In contemporary discussions on gender disparity within the AI industry, the focus has shifted towards examining how knowledge is generated and how cultural norms are perpetuated. Additionally, addressing bias in data sets and algorithms, enhancing transparency in decision-making processes, and promoting diversity are seen as potential solutions to these issues.

This position paper argues that the existing criticisms do not delve deep enough into understanding gendered labor within the AI industry. Additionally, treating the efforts to increase diversity and reduce bias in technology as purely technical optimization challenges is a narrow approach that fails to address the deeper and more complex structural issues at play. It is essential to recognize that despite comprehensive arguments advocating for unbiased and fair AI, ineffective datasets and models persist in use. Furthermore, the industry actively promotes crowd work, which can be exploitative, especially toward marginalized groups \cite{daugherty2018human+}. 

The key insight here is that AI development is not isolated; it operates within the realm of political economy \cite{trajtenberg2018ai}. The profit-driven motives at both the company and state-policy levels drive the AI industry to both benefit from and contribute to the platformization of work. This phenomenon disproportionately disadvantages marginalized workers, particularly women workers \cite{berg2018digital}. This same profit-seeking motive leads the AI industry to increasingly rely on machine learning for hiring and worker surveillance \cite{moore2018humans}. Unfortunately, these practices often disregard limitations such as gender biases \cite{scheuerman2019computers}, the inherent challenges of gender classification via machine learning \cite{hamidi2018gender},  and the oversimplified and non-deterministic nature of these AI artifacts. It also fuels the industry-wide policy apathy towards wage depression, job displacements, and annihilation due to AI and automation - something that disproportionately affects women and the marginalized.

Therefore, it's essential to analyze this issue from a political economy perspective, given its close connection with the increasingly pervasive AI technology \cite{trajtenberg2018ai}. The gender-related impacts of AI are still not thoroughly examined, and ongoing research is unveiling their effects \cite{parsheera2018gendered}. Our position paper seeks to chart uncharted territory by introducing novel lines of investigation in this context.

Machine learning operates by making decisions based on patterns found in extensive historical data \cite{carbonell1983overview}. Consequently, it perpetuates and solidifies the material circumstances of the past. This amplifies preexisting social structures, which include elements of inequality and gender-based oppression. Since machine learning has the ability to identify, reproduce, and even alter both collective and individual behavioral patterns based on the data it learns from, individuals and organizations responsible for developing, designing, and assessing these AI systems should recognize the potential societal impact of these systems and take on a certain level of political responsibility.

\section{Political Economy of AI and Gendered Labour}
The issue of gendered labor being influenced by the AI industry has its roots in material factors, specifically the reliance of technology development on capital ownership \cite{walker1989capitalist,morozov2018rethinking}. To illustrate this, let's consider how material inequality manifested in the period before AI digitization. In 1974, an equal number of men and women expressed interest in pursuing careers in coding. However, between 1981 and 1984, the percentage of women with degrees in computer and informational science dropped significantly to 37.1\% \cite{light1999computers}. Researchers from science and technology studies and gender studies suggest that initially, computing (including programming) was perceived as low-skilled clerical work. As the field gained cultural, social, and, most importantly, economic significance in the early 2000s, a trend emerged in which women were less likely to be hired for these roles \cite{dillon2019ai}. To comprehend the disparities among workers, it is imperative to scrutinize the nature of the work and its intricate relationship with capital.

\subsection{Microwork: neglected labour in AI Industry}
Let's take a closer look at the workforce within the AI industry. It's crucial to recognize that a significant portion of these workers may not have formal recognition but engage in what's known as ``microwork" on distributed platforms \cite{altenried2020platform}. This form of work operates under a piece-wage system and involves workers who are geographically dispersed and working independently.

\begin{figure}[t]
   \centering
   \includegraphics[width=0.9\linewidth]{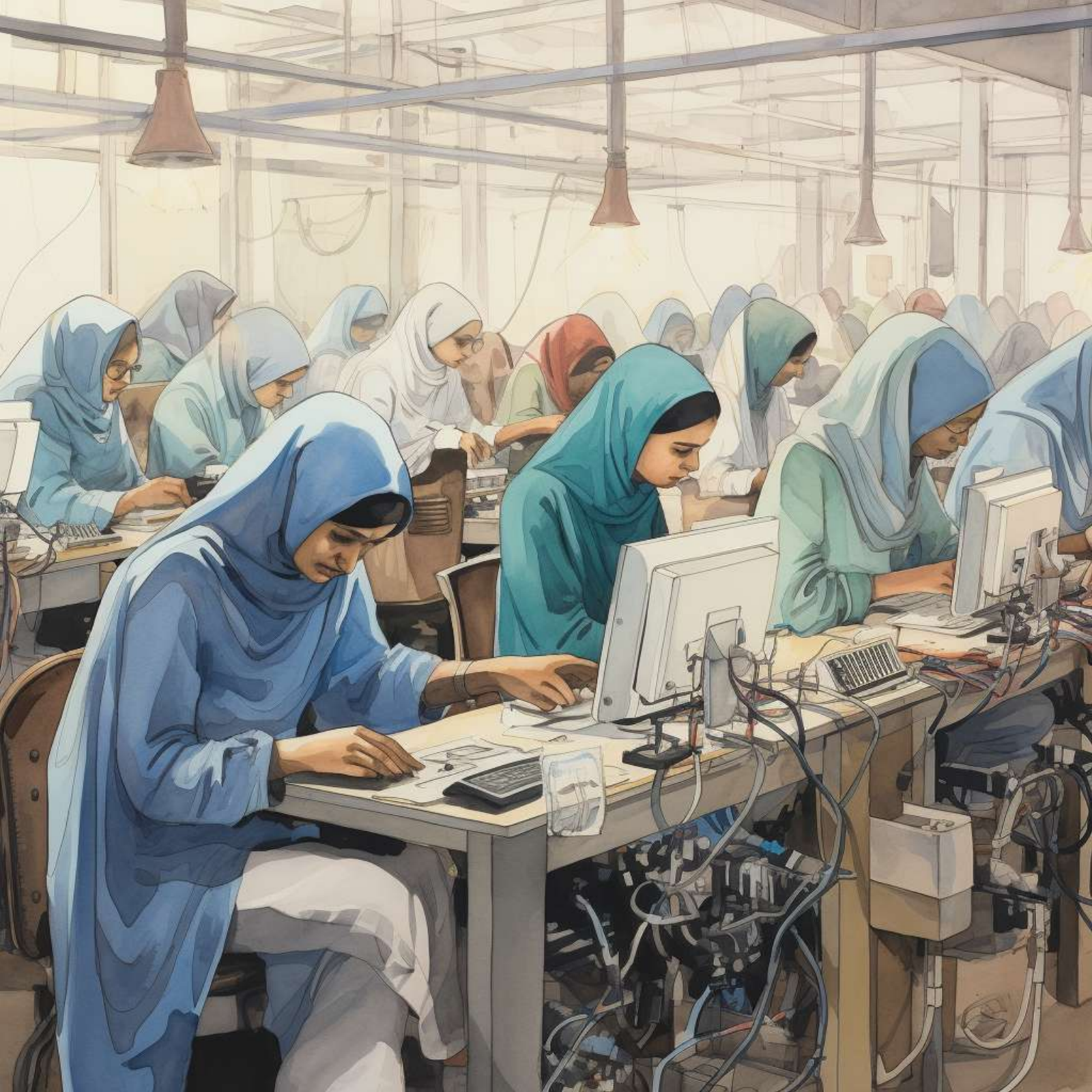}
   \caption{{Microwork done by platform wage workers provides expert} knowledge \& curated data which is essential for machine learning. {However, not only is this work unrecognised, but also often exploitative.}}
   \vspace{-10pt}
\end{figure}

These workers fulfill two crucial roles in the AI industry. Firstly, they are responsible for generating and curating data, which constitutes the largest but often overlooked part of the work essential for machine learning \cite{altenried2020platform}. Secondly, they provide expert knowledge to supervise and moderate machine learning systems. This function is commonly referred to as ``human in the loop," ``artificial artificial systems," ``human intelligence units", etc \cite{link2016human}. Developing this volume of data in-house is often economically unviable, leading to attempts to downplay the extent to which this should be considered "work," along with a global push to relax labor norms in this domain. Consequently, due to the continuous demand for data and human expertise, the AI industry heavily relies on the platformization of work. However, it's worth noting that these platforms themselves significantly depend on the AI industry. They utilize machine learning-centered software (like Uber and Amazon) to operate efficiently and remain competitive. Therefore, the AI industry not only plays a role in promoting the platformization of work but also stands as one of its largest consumers.

Thus, the AI industry plays a significant role in transforming the very nature of work, shifting from traditional waged employment to crowd work on digital platforms. This emerging work model is distinctly exploitative, with profound gender-related consequences. Its exploitative nature arises from several factors. Firstly, it fosters worker atomization and replaceability \cite{irani2015difference}. This is achieved by blurring the boundaries between work and personal life, fragmenting tasks into small, easily replaceable components, erasing distinctions between employees and ``associates," and enabling a form of distributed digital Taylorism \cite{altenried2020platform}. These effects disproportionately affect marginalized individuals, particularly women who often lack representation in unions and similar organizations. Moreover, this new work model discourages women's participation due to the absence of typical employment safeguards, such as gender-specific health benefits and leave policies. When entire industries undergo automation and platformization, the re-skilling process becomes particularly challenging, especially for women who face domestic constraints in patriarchal societies, as they are also responsible for unpaid domestic work. According to a 2017 ILO survey on crowd workers \cite{berg2018digital} 15\% of women cited domestic care work as a factor influencing their work choices, compared to 5\% of men. Thus, for women, crowd work is particularly unattractive, and the transition from conventional employment to digital platforms is more detrimental. 

The growing platformization of work increases the risk of women leaving the workforce entirely, and the AI industry contributes to this trend. The same report notes that only one in every three crowd workers is a woman. A similar gender imbalance is observed among Amazon Turk Workers, CrowdFlower workers, and similar platforms. 

There's a prevailing notion that this kind of work offers liberation, allowing individuals to structure their time and attend to family responsibilities. However, it's important to recognize that due to worker atomization, replaceability, and digital Taylorism, the collective bargaining power of workers\cite{johnston2018organizing,vandaele2018will} has also significantly declined, leading to suppressed wages. These effects are particularly harsh for women workers.

\subsection{The visible AI workers}

Now let's examine the roles within the AI industry that are conventionally recognized. It has been an almost universal phenomenon in tech monopolies that the representation of women in higher-level career positions diminishes \cite{dabla2018women}. This gender imbalance at decision-making levels has repercussions that extend beyond mere representation.

The profit motive drives the AI industry, similar to any other sector, to continually seek cost-cutting measures. This often leads to an increased reliance on AI technologies in areas like hiring and surveillance. Machine learning is increasingly used in the hiring process itself, \cite{liem2018psychology} from screening resumes to the more controversial use of ''emotion detection"\cite{bendel2018uncanny} in interviews. It's important to reiterate a frequently discussed point: these technologies are built on problematic and unscientific assumptions that need to be challenged. Moreover, the uncritical replication of past data through machine learning perpetuates the historical biases of the industry.

One infamous case was Amazon’s attempt at automating the recruitment process. \cite{meyer2018amazon} Amazon assembled a team of engineers in Edinburgh to create an AI tool to economize the hiring process. This tool when developed exhibited indiscriminate bias by rejecting female applicants. This incident laid bare the underlying sexist recruitment practices within the company. It's crucial to recognize that the tool's behavior must have been learned from training data, which was sourced from Amazon's historical practices, thus amplifying an existing biased pattern. Eventually, the tool had to be discontinued.\cite{meyer2018amazon} 

Such arbitrary and biased systems, when directly connected with society, have far-reaching consequences. They hinder gender diversity in the job market, perpetuate the production and dissemination of biased actions and knowledge by AI systems, and influence policy and legal frameworks in countries that regularly interact with AI-driven products in the public sphere.

\begin{figure}[t]
   \centering
   \includegraphics[width=0.8\linewidth]{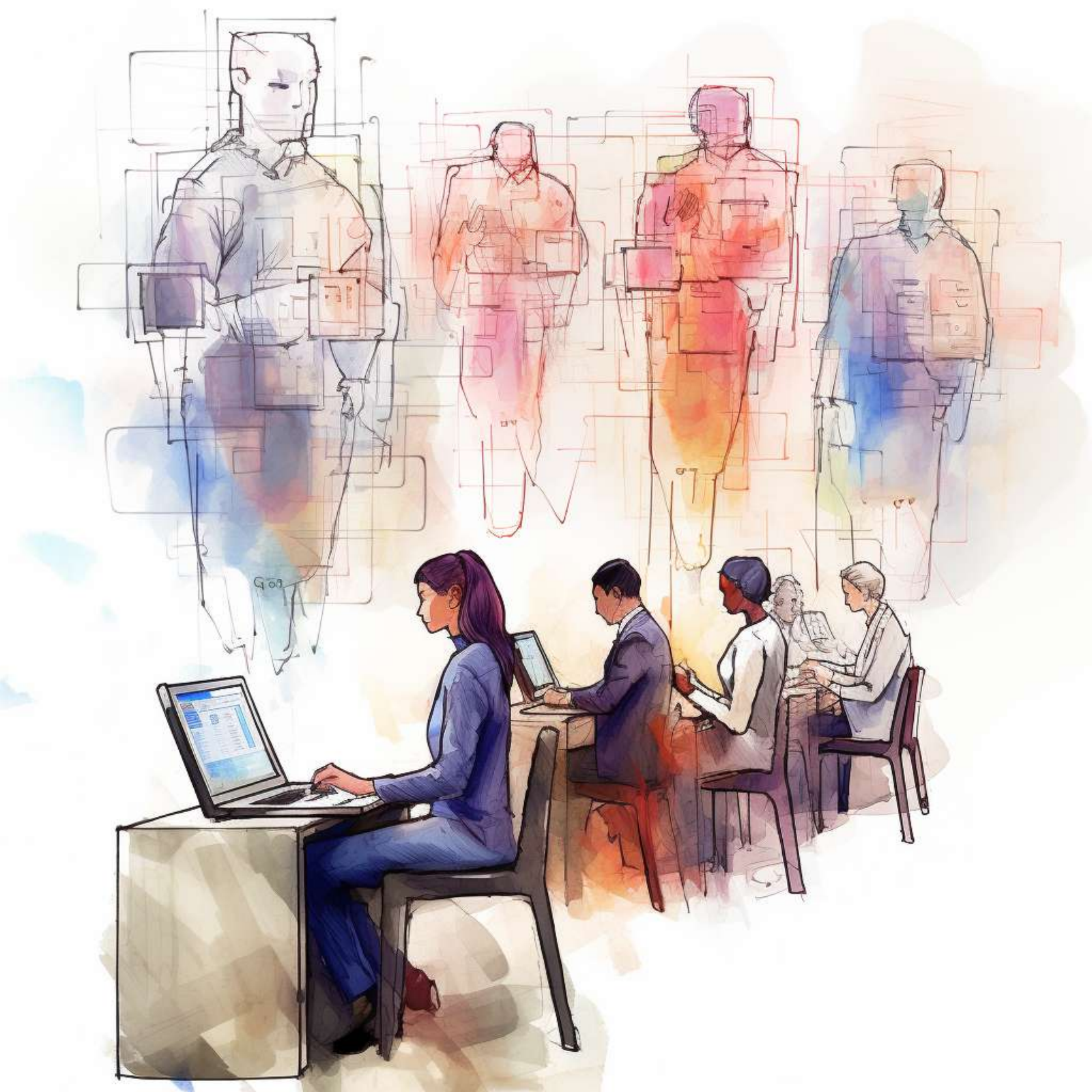}
   \caption{{Assisted recruitment using algorithms for emotion detection,} {filtering etc. has often suffered from biases, including misogyny \& racism.} Emotion algorithms especially promote the pseudoscience of pathognomy, \& rely on the unfounded assumption that it is possible to accurately infer internal emotional states solely from facial cues.}
   \vspace{-10pt}
\end{figure}

Within the AI industry, the challenges extend beyond the hiring phase, as numerous companies now offer AI products for ongoing worker monitoring. Apart from the arguments highlighting these products as disenfranchising, oppressive, and unscientific, there's a significant issue: they disproportionately impact women and hinder their prospects for career advancement. Take, for instance, emotion detection technology, one of these monitoring tools. It essentially resurrects the pseudoscience of pathognomy, \cite{jandl2017writing} relying on the unfounded assumption that one can accurately infer internal emotional states solely from facial expressions. These technologies not only lack scientific validity but also exhibit biases, including misogyny and racism\cite{rhue2018racial}. \cite{barrett2019emotional} suggest that these technologies employ value-neutral terms like ``pattern of facial movements" instead of ``emotional expression".

The use of these arbitrary technologies in the profit-driven digital economy exacerbates the ongoing erosion of bargaining power among AI tech workers, further marginalizing already vulnerable social groups. This raises a crucial question: Are the creators of these technologies aware of these issues and actively working to mitigate them?

\section{Gendered Impact of the AI Industry}

The impact of the AI industry on gender relations extends beyond women tech workers and touches every facet of academia, policy, politics, and society as a whole. One immediate and adverse consequence of a gendered pipeline into the AI industry is its effect on the prospects of women researchers in AI academia. Studies have revealed that only 18\% of authors in leading AI conferences are women, and over 80\% of AI professors identify as cisgender men
\cite{gagne2019global}. Unfortunately, there is no public data available on the representation of transgender individuals and other gender minorities in AI-related professions. The reasons for the under representation of women and LGBTQIA+ individuals in academia are multifaceted. However, addressing this issue requires providing better job prospects in the industry and rectifying the lack of exposure to technological opportunities for young women and marginalized communities.

Beyond the realms of industry and academia, the rapid integration of AI technologies into the economy has led to a phenomenon known as the ``function creep" of AI artifacts \cite{safdar2016function}.  These artifacts are increasingly assuming roles in policymaking, policing \cite{marda2020data}, and governance without significant democratic scrutiny or acknowledgment that machine learning-powered software, replacing human decision-making, often replicates historical patterns of decision-making. We argue that this represents a failure to fully internalize and communicate the nature of machine learning, which excels at replicating the past - an outcome the research community and the broader AI field seek to avoid when it comes to gender-related matters.

The larger issue at hand goes beyond mere bias, which has garnered substantial research attention. It pertains to the interventions of the AI industry and the resulting non-deterministic nature of machine learning in areas that should ideally fall under the domain of deliberate, transparent, and democratic policymaking. While significant challenges, such as Facial Recognition Technology (FRT), are being tackled by AI researchers, often within the context of bias, the impacts of the smaller yet more pervasive presence of machine learning in everyday artifacts — impacts that have the potential to disproportionately affect women, often remain overlooked.

One of the worrisome consequences of the AI industry's influence is not limited to the women actively engaged in its production but extends to the broader labor market, significantly impacting how women workers are treated. The AI industry is rapidly advancing technologies that are poised to bring about wage depression among middle-income workers and, in certain sectors, the outright elimination of entire job categories. Strikingly, there is a noticeable lack of comprehensive literature addressing whether any new jobs will emerge in sufficient numbers to replace those being phased out.

Similarly, while some researchers insist that these technologies can barely replace jobs \cite{grace2018will} and would in the majority be assisting workers \cite{wilson2018collaborative}, the underlying reality is that assisting technologies often translates to fewer opportunities in lower and middle-skilled roles — the very roles that the technology industry often directs women toward. 

IMF’s Staff Discussion Notes \cite{brussevich2018gender} estimate that 11\% of jobs currently held by women are at risk of being eliminated due to AI and other digital technologies. This percentage is higher than the corresponding risk for men. Additionally, women across the world are disproportionately concentrated in clerical, services, and sales positions, which happen to be the easiest to automate. This situation is further exacerbated in countries of the global south, where a pronounced digital gender divide exists in terms of access to digital infrastructure and educational resources that could have helped mitigate this issue to some extent.

\begin{figure}[t]
   \centering
   \includegraphics[width=1.03\linewidth]{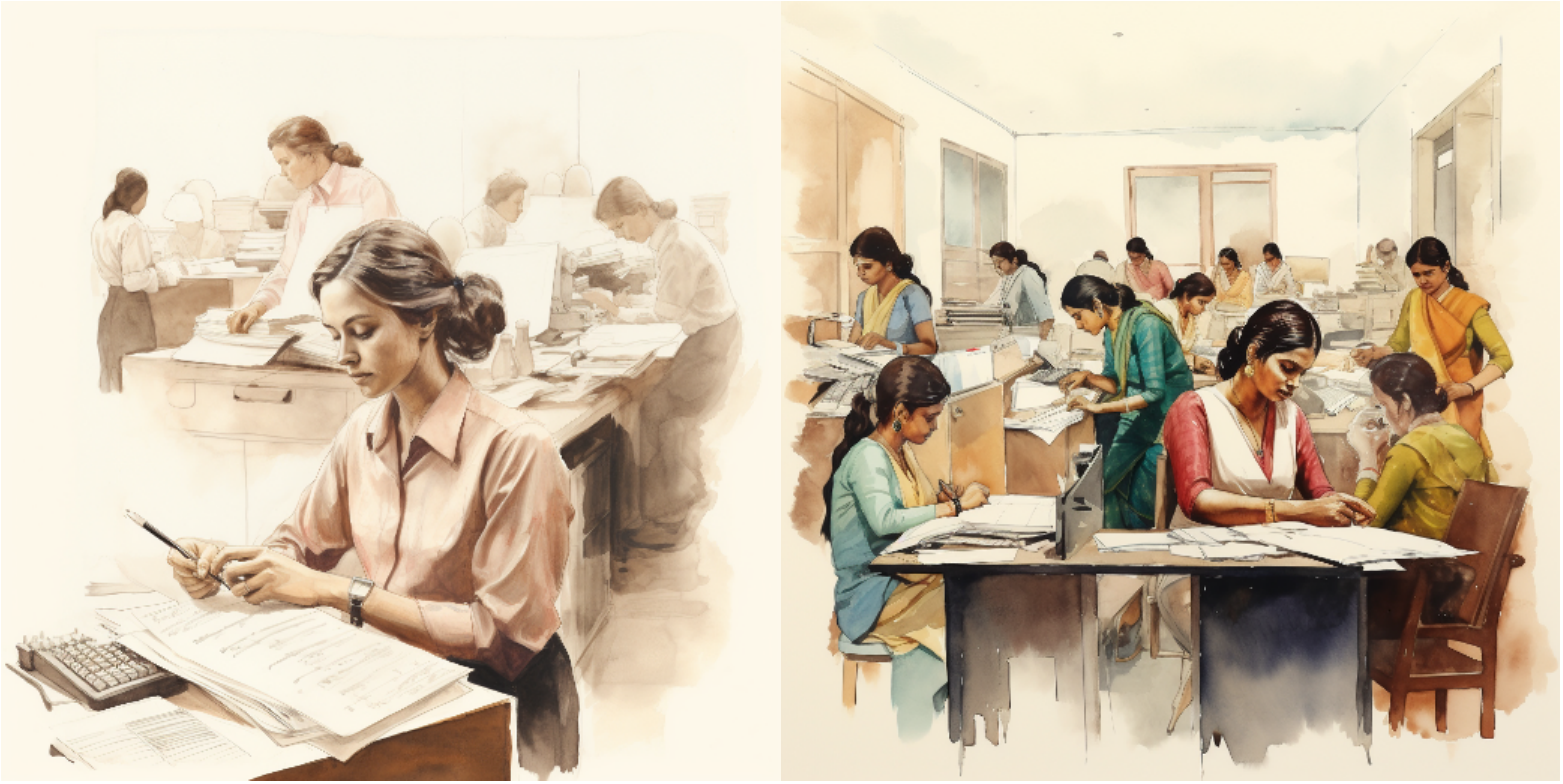}
   \caption{Disproportionate concentration of women in clerical, services, and sales positions put their jobs at risk of being eliminated due to AI. Digital divide, limited digital infrastructure and lack of access to education exacerbates the issue further.}
    \vspace{-10pt}
\end{figure}

The presence of AI technologies, automating some jobs and drastically transforming others, often necessitates job transitions for workers. For women, this transition can be more challenging due to the burden of domestic responsibilities and the high monetary and time costs associated with reskilling. This challenge is particularly pronounced in the global south, where the absence of public healthcare, social security, and gender-specific benefits compounds the difficulties faced by women in remaining in the workforce.

% \begin{figure}
%   \includegraphics[width=\linewidth]{graph.jpg}
%   \caption{stats for jobs US data graph}
%   \label{fig:graph}
% \end{figure}

\section{What is to be done?}
In this position paper, we have advocated for adopting a perspective rooted in political economy to assess the AI industry's impact on gendered labor and the resulting gender-based oppression. The AI industry is not merely a collection of technologies but represents a profound transformation in the processes of production, distribution, and exchange within the framework of capitalism. Consequently, we have explored how women workers are positioned within this industry and its broader ramifications. The AI industry's scope extends far beyond what is typically imagined, as it is intricately intertwined with various platforms, and its effects spill over into the labor market, knowledge production, and policymaking. Its influence on women transcends biased decision-making, and we argue that as AI researchers if we are genuinely concerned about the societal consequences of our creations, our focus should go beyond merely scrutinizing our artifacts for bias and transparency. We should also delve into analyzing how these artifacts reshape social and economic relationships.

While this paper does not provide an exhaustive exploration of the topic, its purpose is to inspire a research direction grounded in the political economy of gendered labor, job displacement, and the gender-specific consequences of AI-driven policymaking. It is crucial to stress that studying AI artifacts in isolation is insufficient for understanding their social implications, especially concerning gender. In this paper, we have put forth a framework for future research and raised questions aimed at fostering ongoing discussions and investigations in this critical area.

%%%%%%%%%%%%%%%%%%%%%%%%%%%%%%%%%%%%%%%%%%%%%%%%%%%%%%%%%%%%%%%%%%%%%%%%%%%%%%%%%%%%%%%%%%%%%%%%%%%%%%%%%%%%%%%%%%%%%%%%%%%%%%%%%%%%%%%%%%%%%%%%%%%%%%%%%%%%%%%%%%%%%%%%%%%%%%%%%%%%%%%%%%%%%%%%%%%%%%%%%%%%%%%%%%%%%%%%%%%%%%%%%%%%%%%%%%%%%%%%%%%%%%%%%%%%%%%%%%%%%%%%%%%%%%%%%%%%%%%%%%%%%%%%%%%%%%%%%%%%%%%%%%%%%%%%%%%%%%%%%%%%%%%%%%%%%%%%%%%%%%%%%%%%%%%%%%%%%%%%%%%%%%%%%%%%%%%%%%%%%%%%%%%%%%%%%%%%%%%%%%%%%%%%%%%%%%%%%%%%%%%%%%%%%%%%%%%%%%%%%%%%%%%%%%%%%%%%%%%%%%%%%%%%%%%%%%%%%%%%%%%%%%%%%%%%%%%%%%%%%%%%%%%%%%%%%%%%%%%%%%%%%%%%%%%%%%%%%%%%%%%%%%%%%%%%%%%%%%%%%%%%%%%%%%%%%%%%%%%%%%%%%%%%%%%%%%%%%%%%%%%%%%%%%%%%%%%%%%%%%%%%%%%%%%%%%%%%%%%%%%%%%%%%%%%%%%%%%%%%%%%%%%%%%%%%%%%%%%%%%%%%%%%%%%%%%%%%%%%%%%%%%%%%%%%%%%%%%%%%%%%%%%%%%%%%%%%%%%%%%%%%%%%%%%%%%%%%%%%%%%%%%%%%%%%

\bibliographystyle{ACM-Reference-Format}
\bibliography{sample-base.bib}

\end{document}